\documentclass[pss]{wiley2sp} 
\usepackage{amsmath}

\begin{document}

\title{Semiconductors for Plasmonics and Metamaterials}

\titlerunning{Semiconductor Plasmonic Metamaterials}

\author{%
  Gururaj V. Naik\textsuperscript{\textsf{\bfseries 1}},
  Alexandra Boltasseva\textsuperscript{\Ast,\textsf{\bfseries 1,2}}}

\authorrunning{G. V. Naik et al.}

\mail{e-mail
  \textsf{aeb@purdue.edu}, Phone:
  +1-765-4940301, Fax: +1-765-4946951}

\institute{%
  \textsuperscript{1}\, School of Electrical \& Computer Engineering and Birck Nanotechnology Center, Purdue University IN 47907-2057 USA\\
  \textsuperscript{2}\, DTU Fotonik, Technical University of Denmark, Lyngby, DK-2800, Denmark}

\published{at http://onlinelibrary.wiley.com/doi/10.1002/pssr.201004269/abstract} 


\abstract{%
\abstcol{%
  Plasmonics has conventionally been in the realm of metal-optics. However, conventional metals as plasmonic elements in the near-infrared (NIR) and visible spectral ranges suffer from problems such as large losses and incompatibility with semiconductor technology. Replacing metals with semiconductors can alleviate these problems if only semiconductors could exhibit negative real permittivity. Aluminum doped zinc oxide (AZO) is 
  }{%
   a low loss semiconductor that can show negative real permittivity in the NIR. A comparative assessment of AZO-based plasmonic devices such as superlens and hyperlens with their metal-based counterparts shows that AZO-based devices significantly outperform at a wavelength of 1.55 $\mu m$. This provides a strong stimulus in turning to semiconductor plasmonics at the telecommunication wavelengths.}}

\maketitle   

\section{Introduction}
Plasmonics as a route to sub-wavelength optics has aroused the curiosity of a variety of researchers. Its numerous applications range from sensors and imaging to waveguides and switches \cite{plasmonics_ozbay}. The vast potential of plasmonics can be better exploited if the losses in the plasmonic components are reduced and the integration with semiconductor technology is feasible \cite{plasmonics_brongersma}. Both of these conditions can be met if plasmonic materials can be semiconductors instead of conventional metals such as gold and silver. Unlike metals, heavily doped semiconductors can exhibit a small negative real permittivity ($\epsilon'$) and very small losses at infrared and longer wavelengths \cite{negref_hoffman}. However, achieving a negative $\epsilon'$ in the NIR is non-trivial because extremely high doping would be necessary. In this paper, we outline the development of a low-loss semiconductor plasmonic material for the NIR and present the design of novel plasmonic devices such as superlens and hyperlens using this material at a wavelength of 1.55 $\mu m$. We also provide a quantitative assessment of the performance of these semiconductor plasmonic systems at 1.55 $\mu m$ as against the conventional metal based systems.

\section{Doped semiconductors}
Optical properties of heavily doped semiconductors can be described by Drude model \cite{drude_ashcroft} in the absence of interband transitions. Accordingly, negative $\epsilon'$ in semiconductors at optical frequencies may be achieved by large doping and small $\epsilon_{int}$ (low frequency permittivity of semiconductor with no free carriers) and, lower losses by small damping and absence of interband transitions. Smaller damping implies large carrier mobility and, avoiding interband transitions in the spectrum of interest implies bandgap to be larger than the largest photon energy of interest. With these guidelines, we next assess various semiconductors and intermetallic nitrides for their suitability as NIR plasmonic materials.

\subsection{Semiconductors plasmonic materials}
Conventional semiconductors such as silicon and germanium, III-Vs such as aluminum, gallium and indium phosphides, arsenides and antimonides may be grown as single crystals with very high carrier mobilities. However, the primary problems with these semiconductors are their large $\epsilon_{int}$ and difficulty in doping beyond $10^{20}cm^{-3}$ due to the electrical solubility limit of dopants \cite{plummer2000silicon,schubert1993doping}.

Heavy doping of about $10^{21} cm^{-3}$ is possible in oxide semiconductors such as zinc oxide and indium oxide. The solid-solubilities of dopants (Al, Ga in ZnO and Sn in $In_2O_3$) are very high which makes large doping possible. These semiconductors have wide band-gaps and their $\epsilon_{int}$ values are moderately low. In addition, relatively high mobilities of charge carriers can be achieved with polycrystalline microstructure though very high mobilities are typically observed in single crystals. Thus, these semiconductors are good candidates for plasmonic applications \cite{reviewTCO_minami}.

\begin{figure}[tb]%
\includegraphics*[width=0.95\linewidth,height=0.72\linewidth]{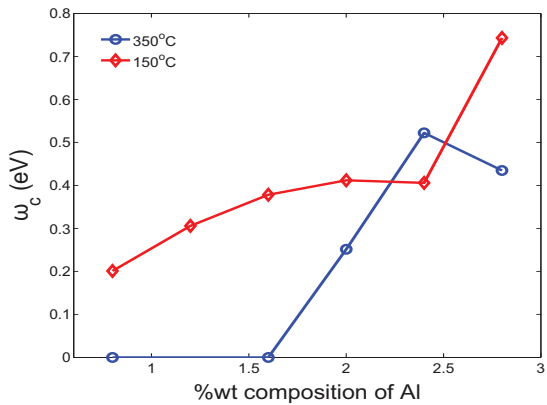}
\caption{Optimization plot for pulsed laser deposited (PLD) aluminum doped zinc oxide (AZO) thin films showing the cross-over frequency ($\omega_c$) as a function of doping concentration for two different deposition temperatures.}
\label{fig1}
\end{figure}

In reference \cite{APM_LPR} we showed that aluminum doped zinc oxide (AZO) can have losses as low as four-times smaller than that of silver in the NIR and, Gallium doped Zinc Oxide (GZO) and Indium-Tin-Oxide (ITO) are also low-loss plasmonic materials in the NIR. The properties of these oxide semiconductors are strongly dependant on their fabrication procedures. When optimized however, AZO shows the lowest losses \cite{APM_LPR}. Pulsed-laser-deposition (PLD) which is reported to produce films with high carrier mobility is adopted to deposit thin films of AZO \cite{reviewTCO_minami}. Optimizing doping is a challenge as too high doping can result in low donor-ionization-efficiency and solid-solubility problems \cite{SolidSolubility_yoon}. The deposition temperature and oxygen partial pressure during deposition play significant roles in the optimization. Figure \ref{fig1} shows the optimization curves for AZO films. The films were characterized by variable-angle spectroscopic ellipsometer (J.A. Woollam Co.) to extract the Drude parameters. The smallest cross-over wavelength (wavelength at which $\epsilon'$ crosses zero) observed so far is 1720 nm. The dependence of cross-over frequency on doping shown in Fig. \ref{fig1} suggests that further increase in doping could increase the cross-over frequency. In reference \cite{APM_LPR}, we notice that cross-over wavelength as low as 1.3 $\mu m$ is possible with AZO.

Among the intermetallics such as silicides and germanides, nitrides exhibit lower losses and larger carrier concentrations. The interband transitions, however, are not completely absent in the NIR and visible spectral ranges making them less attractive as plasmonic materials. However, unlike metals, their $\epsilon'$ values can be small in magnitude and negative in visible spectrum, which make them potential plasmonic materials for niche applications. Titanium nitride is a transparent nitride exhibiting negative $\epsilon'$ for wavelengths longer than about 500 nm \cite{TiN_patsalas}. With some optimization, the losses in titanium nitride can be reduced low enough to be useful for plasmonic applications.

\section{Semiconductor plasmonic devices}
Semiconductor plasmonics in the NIR has been demonstrated with ITO for experiments involving the excitation of SPPs \cite{ITO_franzen}. However, there has not been much progress beyond that as semiconductors cannot have large negative $\epsilon'$ in the NIR. For plasmonic waveguide applications, semiconductors with small negative $\epsilon'$ provide much better confinement than metals \cite{book_maier}. For example, while the 1/e field spread of a single-interface AZO/air waveguide can be about 350 nm at a wavelength of 1.55 $\mu m$ , the same for a gold/air interface is about 2.5 $\mu m$. However, this comes at the cost of propagation length which is reduced from about 200 $\mu m$ for gold to about 4 $\mu m$ for AZO. In comparison, the long ranging-SPP gap mode \cite{book_maier} in a gold/air/gold waveguide with a 100 nm air gap has 1/e field spread of about 270 nm with a propagation length of about 10 $\mu m$ at 1.55 $\mu m$ wavelength \cite{wvgds_optimization_brongersma}. Though the performance of AZO is lower in terms of propagation length, AZO can be effective for niche applications such as plasmonic chemical sensors.

 Semiconductors have unbeatable performance for novel plasmonic devices such as hyperlens and superlens as they have small negative $\epsilon'$. As evidence, the demonstration of negative refraction in a hyperbolic metamaterial (HMM) made of GaAs/AlGaAs layers operating at a wavelength of 9 $\mu m$ had a figure-of-merit (FOM) as high as about 40 \cite{negref_hoffman}. If the same structure is made of silver and dielectric alternating layers, it would have had a FOM close to zero all over the mid-IR and NIR ranges. With metals the FOM improves only near their cross-over frequency if inter-band transitions are absent. This condition is only barely met by conventional metals and, silver being the best candidate so far, has been demonstrated to work well for hyperlens only in UV at 365 nm \cite{hyperlens_zhang}. Similarly, superlens has also been demonstrated only in UV \cite{superlens_zhang}.

\begin{figure}[htb]%
\includegraphics*[width=\linewidth,height=0.72\linewidth]{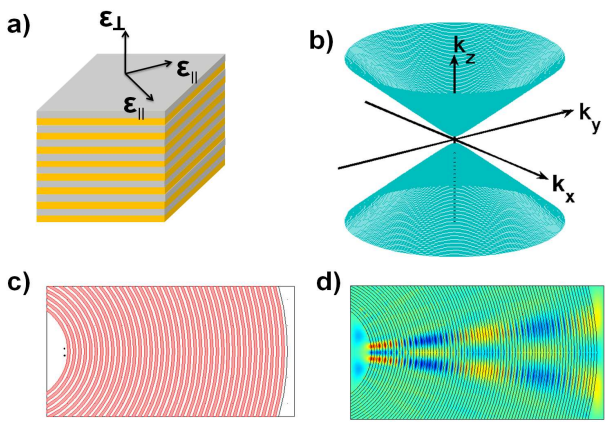}
\caption{a) Schematic of hyperbolic metamaterial (HMM) formed by alternating layers of metal and dielectric b) Dispersion plot of HMM c) Truncated schematic of cylindrical hyperlens formed by alternating layers of metal and dielectric. d) Simulation results plotting the field distribution in a hyperlens demonstrating its capability to resolve two point sources separated by $\lambda/4$ in the far-field.}
\label{fig2}
\end{figure}

In order to develop an insight into this problem, consider a schematic of a HMM formed by alternating layers of metal and dielectric as shown in Fig. \ref{fig2}a. If the individual layers are much thinner than the wavelength, from effective medium approximation (EMA) the anisotropic permittivity values are given by Eq.~\ref{hmm}, where $\epsilon_m$ and $\epsilon_d$ are permittivity values of metal and dielectric respectively and \emph{f} is the metal filling fraction.

\begin{eqnarray}
\label{hmm}
\epsilon_{\parallel}=f\epsilon_m+(1-f)\epsilon_d & \textbf{,} & \epsilon_{\perp}=\frac{\epsilon_m\epsilon_d}{f\epsilon_d+(1-f)\epsilon_m}
\end{eqnarray}

Hyperbolic dispersion arises when real parts of $\epsilon_{\parallel}$ and $\epsilon_{\perp}$ are of opposite signs. The k-space of HMM would be as shown in Fig. \ref{fig2}b. Unlike the spherical dispersion of vacuum, HMM allows infinitely large propagating wavevectors. This principle is utilized in many devices such as the hyperlens to achieve sub-wavelength resolution imaging \cite{book_wenshen}. For example, the HMM structure can be a multi-layered equivalent of a single slab superlens \cite{superlens_ramakrishna} if $Re\{\epsilon_{\parallel}\}=0$. Using this in Eq.~\ref{hmm} gives $f=\frac{\epsilon_d}{\epsilon_d-Re\{\epsilon_m\}}$. At a wavelength of 1.55 $\mu m$,  even for high permittivity dielectric such as silicon, the silver filling fraction turns out to be less than 10\%. Such low filling fraction requires ultra-thin silver films close to the percolation limit. The FOM \cite{APM_LPR,superlens_ramakrishna} would be 0.25 without considering the additional losses that would arise from the roughness of each layer and the associated scattering. All of these problems can be overcome with nearly the same FOM for a single-slab AZO superlens.

For a negatively refracting HMM slab, the FOM may be defined as $FOM=Re\{k_{\perp}\}/Im\{k_{\perp}\}$ \cite{negref_hoffman}. FOM is greater than unity if only $Re\{\epsilon_{\parallel}\}>0$. It may be noticed from Eq.~\ref{hmm} that no metal filling fraction can satisfy this condition for any of the conventional metals in the NIR. On the contrary, if AZO/ZnO layer stack with 50\% filling fraction is used as a HMM, the FOM can be as high as 5.5 for a polycrystalline structure and about 65 for a single crystal system. The AZO/ZnO HMM stack if formed as cylindrical layers instead of planar layers, performs as hyperlens. The geometry of hyperlens is schematically shown in Fig. \ref{fig2}c together with the simulation-based illustration of achievable sub-wavelength resolution (see Fig. \ref{fig2}d). Using AZO/ZnO to build such as device at NIR would result in FOM of about 5.5. As a comparison, the Ag/$Al_2O_3$ system used to demonstrate hyperlens at 365 nm wavelength \cite{hyperlens_zhang} had FOM of about 1.5. Thus, AZO significantly outperforms any conventional metal for the above mentioned metamaterial applications. As a recent development, AZO/ZnO HMM multilayer stack is a good candidate for developing quantum optics devices based on engineering the photonic density of states \cite{PDOSengg_jacob}. Thus, AZO is a good choice as a semiconductor plasmonic material in the NIR for the novel plasmonic and metamaterial-based devices for nanophotonics and quantum optics.

\section{Conclusions}
Low loss, heavily doped semiconductors as plasmonic materials offer many advantages over conventional metals for nanophotonic applications in the NIR. They facilitate the integration of plasmonics and nanophotonics into the well-established semiconductor technology and thereby stimulating the development of new types of nano-optical devices \cite{additional}.

\begin{acknowledgement}
We gratefully acknowledge the support and guidance of Prof. Vladimir M. Shalaev and thank Zubin Jacob for helpful discussions. This work is supported by ARO grant W911NF-09-1-0516.
\end{acknowledgement}

\providecommand{\WileyBibTextsc}{}
\let\textsc\WileyBibTextsc
\providecommand{\othercit}{}
\providecommand{\jr}[1]{#1}
\providecommand{\etal}{~et~al.}


\begin{thebibliography}{[10]}

\bibitem{plasmonics_ozbay}
 \textsc{E.~Ozbay},
 \jr{Science} \textbf{311}(5758), 189 (2006).


\bibitem{plasmonics_brongersma}
 \textsc{M.~Brongersma},  \textsc{R.~Zia},  and  \textsc{J.~Schuller},
 \jr{Applied Physics A: Materials Science \& Processing} \textbf{89}(2),
  221--223 (2007).


\bibitem{negref_hoffman}
 \textsc{A.~Hoffman},  \textsc{L.~Alekseyev},  \textsc{S.~Howard},
  \textsc{K.~Franz},  \textsc{D.~Wasserman},  \textsc{V.~Podolskiy},
  \textsc{E.~Narimanov},  \textsc{D.~Sivco},  and  \textsc{C.~Gmachl},
 \jr{Nature Materials} \textbf{6}(12), 946--950 (2007).


\othercit
\bibitem{drude_ashcroft}
 \textsc{N.~Ashcroft} and  \textsc{N.~Mermin},
{Solid Stale Physics} (Saunders College, Philadelphia, 1976).


\othercit
\bibitem{plummer2000silicon}
 \textsc{J.~Plummer},  \textsc{M.~Deal},  and  \textsc{P.~Griffin},
{Silicon VLSI technology} (Prentice Hall Upper Saddle River, NJ, 2000).


\othercit
\bibitem{schubert1993doping}
 \textsc{E.~Schubert},
{Doping in III-V semiconductors} (Cambridge University Press New York, 1993).


\bibitem{reviewTCO_minami}
 \textsc{T.~Minami},
 \jr{MRS Bulletin} \textbf{25}(8), 38--44 (2000).


\bibitem{APM_LPR}
 \textsc{P.~West},  \textsc{S.~Ishii},  \textsc{G.~Naik},  \textsc{N.~Emani},
  \textsc{V.~Shalaev},  and  \textsc{A.~Boltasseva},
 \jr{Laser \& Photonics Reviews, DOI:10.1002/lpor.200900055} p.\,NA (2010).


\bibitem{SolidSolubility_yoon}
 \textsc{M.~Yoon},  \textsc{S.~Lee},  \textsc{H.~Park},  \textsc{H.~Kim},  and
  \textsc{M.~Jang},
 \jr{Journal of Materials Science Letters} \textbf{21}(21), 1703--1704 (2002).


\bibitem{TiN_patsalas}
 \textsc{P.~Patsalas} and  \textsc{S.~Logothetidis},
 \jr{Journal of Applied Physics} \textbf{90}(9), 4725--4734 (2001).


\bibitem{ITO_franzen}
 \textsc{S.~Franzen},
 \jr{The Journal of Physical Chemistry C} \textbf{112}(15), 6027--6032 (2008).


\othercit
\bibitem{book_maier}
 \textsc{S.~Maier},
{Plasmonics: fundamentals and applications} (Springer Verlag, 2007).


\bibitem{wvgds_optimization_brongersma}
 \textsc{R.~Zia},  \textsc{M.~Selker},  \textsc{P.~Catrysse},  and
  \textsc{M.~Brongersma},
 \jr{Journal of the Optical Society of America A} \textbf{21}(12), 2442--2446
  (2004).


\bibitem{hyperlens_zhang}
 \textsc{Z.~Liu},  \textsc{H.~Lee},  \textsc{Y.~Xiong},  \textsc{C.~Sun},  and
  \textsc{X.~Zhang},
 \jr{Science} \textbf{315}(5819), 1686 (2007).


\bibitem{superlens_zhang}
 \textsc{N.~Fang},  \textsc{H.~Lee},  \textsc{C.~Sun},  and
  \textsc{X.~Zhang},
 \jr{Science} \textbf{308}(5721), 534 (2005).


\othercit
\bibitem{book_wenshen}
 \textsc{W.~Cai} and  \textsc{V.~Shalaev},
{Optical Metamaterials: Fundamentals and Applications} (Springer Verlag, 2009).


\bibitem{superlens_ramakrishna}
 \textsc{S.~Ramakrishna},  \textsc{J.~Pendry},  \textsc{M.~Wiltshire},  and
  \textsc{W.~Stewart},
 \jr{Journal of Modern Optics} \textbf{50}(9), 1419--1430 (2003).


\bibitem{PDOSengg_jacob}
 \textsc{Z.~Jacob},  \textsc{J.~Kim},  \textsc{G.~Naik},
  \textsc{A.~Boltasseva},  \textsc{E.~Narimanov},  and
  \textsc{V.~Shalaev},
 \jr{Arxiv preprint arXiv:1005.5172} (2010).
 
 \bibitem{additional}
 \textsc{This paper is published as:} \textsc{G.~Naik},  \textsc{A.~Boltasseva} ,
 \jr{physica status solidi -(RRL) Rapid Research Letters} \textbf{4}(10), 295--297 (2010).


\end{thebibliography}

\end{document}